\documentclass[prd,letter,nofootinbib
,twocolumn,preprintnumbers,showpacs]{revtex4}
\usepackage{subfigure,feynmp,graphicx,amsfonts,color,comment,amsmath,hyperref}
\usepackage{times}
\usepackage[T1]{fontenc}\usepackage{textcomp}\usepackage{mathptmx}

\usepackage{amssymb}

\def\hbar{\hspace{0pt}\raisebox{1pt}{$-$} \hspace{-7pt} h}

\def\5{\overline 5}

\newcommand{\be}{\begin{equation}}
\newcommand{\ee}{\end{equation}}
\newcommand{\bea}{\begin{eqnarray}}
\newcommand{\eea}{\end{eqnarray}}

\newcommand{\ba}{\begin{eqnarray}}
\newcommand{\ea}{\end{eqnarray}}

\newcommand{\nova}{NO$\nu$A}
\newcommand{\epset}{\varepsilon_{e\tau}}

\begin{document}
\title{Searching for Novel Neutrino Interactions at NO$\nu$A and Beyond in Light of Large $\theta_{13}$}

\author{Alexander Friedland}
\email{friedland@lanl.gov}
\author{Ian M. Shoemaker}
\email{ianshoe@lanl.gov}
\affiliation{Theoretical Division T-2, MS B285, Los Alamos National Laboratory, Los Alamos, NM 87545, USA}

\date{July 27, 2012}

\begin{abstract}

We examine the prospects of probing nonstandard interactions (NSI) of neutrinos in the $e-\tau$ sector with upcoming long-baseline $\nu_{\mu} \rightarrow \nu_{e}$ oscillation experiments. First conjectured decades ago, neutrino NSI remain of great interest, especially in light of the recent $^{8}B$ solar neutrino measurements by SNO, Super-Kamiokande, and Borexino.  We observe that the recent discovery of large $\theta_{13}$ implies that long-baseline experiments have considerable NSI sensitivity, thanks to the interference of the standard and new physics conversion amplitudes. 
In particular, in some parts of NSI parameter space, the upcoming NO$\nu$A experiment will be sensitive enough to see $\sim3\sigma$ deviations from the SM-only hypothesis. On the flip side, NSI introduce important ambiguities in interpreting NO$\nu$A results as measurements of $CP$-violation, the mass hierarchy and the octant of $\theta_{23}$. In particular, observed $CP$ violation could be due to a phase coming from NSI, rather than the vacuum Hamiltonian.  The proposed LBNE experiment, with its longer $\sim 1300 $ km baseline, may break many of these interpretative degeneracies.

 \end{abstract}
\preprint{LA-UR-12-22243}
\maketitle

\section{Introduction}
\label{sec:intro}

\begin{quote}
``\emph{The effect of coherent forward scattering must be taken into account when considering the oscillations of neutrinos traveling through matter. In particular $\left[ \dots \right] $  oscillations can occur in matter if the neutral current has an off-diagonal piece connecting different neutrino types. Applications discussed are solar neutrinos and a proposed experiment involving transmission of neutrinos through 1000 km of rock.}" 
\end{quote}
 
Though the above quote could easily have been written this year, or even applied to the present paper, it was written presciently  in 1978 by Lincoln Wolfenstein in his seminal paper on the effects of matter on neutrino oscilations~\cite{Wolfenstein:1977ue}.  Although originally proposed as an alternative to mass induced oscillations~\cite{Wolfenstein:1977ue,Valle:1987gv,Guzzo:1991hi,Roulet:1991sm}, beyond-the-Standard-Model (BSM) neutrino-quark interactions remain a phenomenological possibility ({\it e.g.}, \cite{Grossman:1995wx,Fornengo:2001pm,Guzzo:2001mi,Davidson:2003ha,Friedland:2004pp,Alex2,Friedland:2005vy,Friedland:2006pi,Antusch:2008tz,Gavela:2008ra,Escrihuela:2009up,globalNSI,Maltoni2011}) that can produce potentially observable effects in oscillation experiments.  Three decades after the above quote was written, we have finally reached the era of 1000 km experiments, with several years of data collected at MINOS, \nova~launching next year, and LBNE on the drawing board. 

Our goal in this paper is to gauge the sensitivity of these experiments to NSI, in light of what has become known about neutrino oscillations over the last decade. We deliberately choose to avoid a full analysis that scans over many couplings with different flavor combinations and consider a  simplified framework with only one effective flavor off-diagonal piece connecting electron- and tau-type neutrinos, $\mathcal{L} \supset \
-2 \sqrt{2}~ \varepsilon_{e\tau}^{f} G_{F}~ \left(\overline{f} \gamma_{\mu}  f \overline{\nu}_{e} \gamma^{\mu} \nu_{\tau}\right) + h.c.$, where $f= u,d,e$.  We will see that this framework nonetheless reveals a rich spectrum of physical possibilities. Importantly, $\epset$ has its own $CP$-violating phase and can lead to ambiguity in interpreting the searches of $CP$-violation and the mass hierarchy.


\begin{figure}[b] 
\begin{center}
\includegraphics[width=\columnwidth]{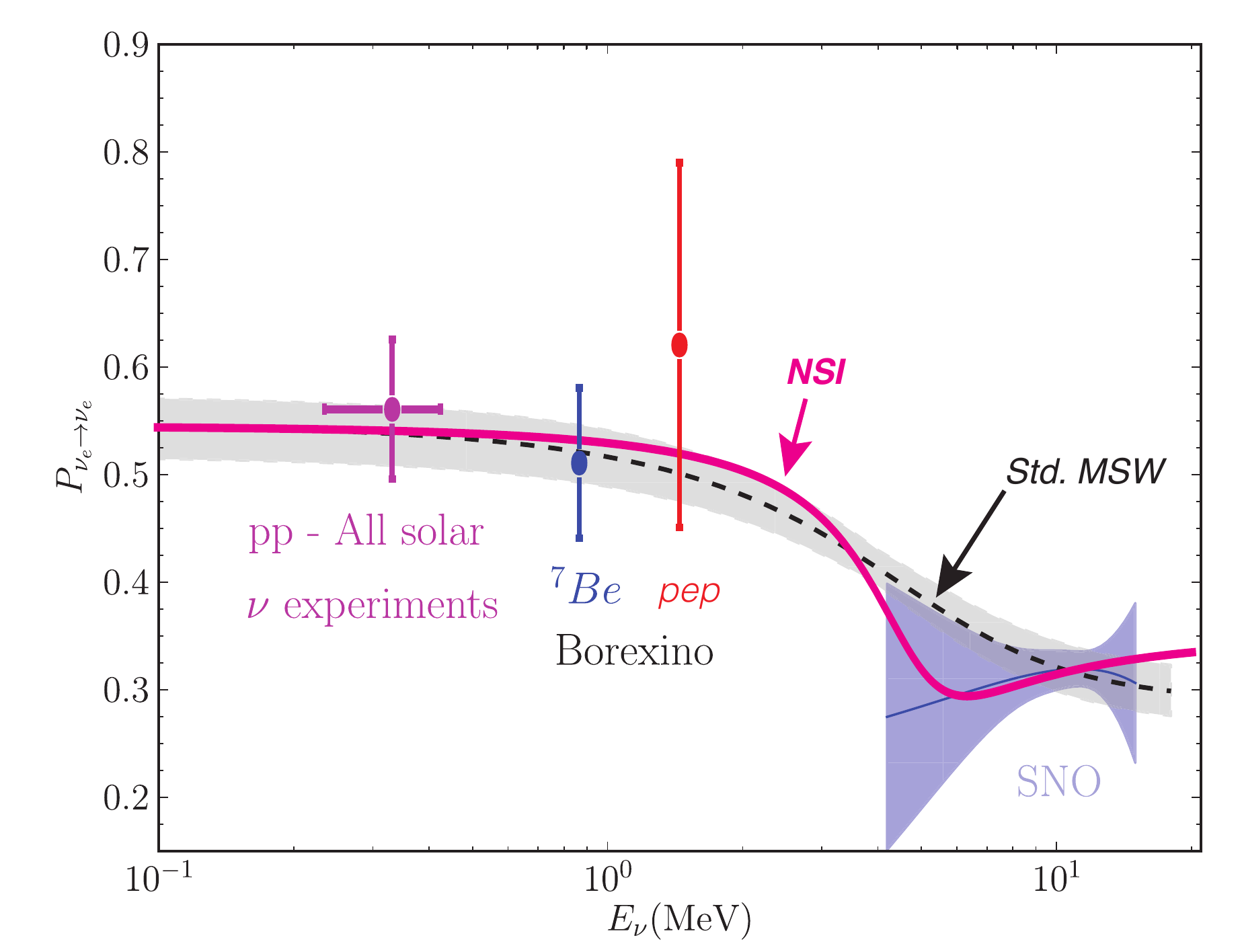}
\caption{\small  Recent SNO solar neutrino data~\cite{SNO2011} on $P(\nu_{e} \rightarrow \nu_{e})$ (blue line with 1 $\sigma$ band). The LMA MSW solution (dashed black curve with gray 1 $\sigma$ band) appears divergent around a few MeV, whereas for NSI with $\varepsilon_{e \tau}=0.4$ (thick magenta), the electron neutrino probability appears to fit the data better. The data points come from the recent Borexino paper~\cite{borexino2}.
\label{fig:SNO}}
\end{center}
\end{figure}

As a first illustration, let us examine the effect this one parameter can have on the solar electron neutrino survival probability, $P(\nu_{e} \rightarrow \nu_{e})$.  The standard large mixing angle (LMA) MSW solution makes a definite prediction for how this probability varies as a function of the neutrino energy, $E_{\nu}$.  This prediction is shown as a dashed line in Fig.~\ref{fig:SNO}, with the gray band around it coming from the uncertainty on the standard oscillation parameters. Both are taken from~\cite{SNO2011}.  Also taken from~\cite{SNO2011} is the allowed region of this probability inferred from all three stages of SNO data, as labeled in Fig.~\ref{fig:SNO}. At low energies, we also include the survival probabilities of $pp$ and $^7Be$ neutrinos and the $pep$ neutrinos as recently reported by Borexino~\cite{borexino2}.  While the standard MSW solution provides an acceptable fit to all this data, the addition of the $\epset$ NSI parameter noticeably improves the fit, as shown with the thick magenta curve.  This finding is consistent with earlier work~\cite{Friedland:2004pp,Palazzo}, although notice that here we vary only the off-diagonal $e-\tau$ coupling, consistent with our framework.  Note that although we have chosen $\varepsilon_{e\tau} = 0.4$ in Fig.~\ref{fig:SNO} (see~\ref{NSIeqs} for the definition of $\epset$), there is a family of continuous deformations that approach the MSW LMA solution as $\varepsilon_{e\tau}$ goes to zero.  The NSI analysis in~\cite{Palazzo}, which preceded the three-stage SNO data release~\cite{SNO2011}, found a best fit of $\epset =0.2$. We hence focus on $|\epset|$ of the order $0.2-0.4$ as fiducial values.

Importantly, this choice is consistent with the wealth of other experimental information. For example, atmospheric neutrino data from Super-Kamiokande allow values of $|\epset| \lesssim 0.5$~\cite{Alex2,Friedland:2005vy,Friedland:2006pi,Maltoni2011}, and become even less restrictive if tunings between different NSI parameters~\cite{Friedland:2005vy} are allowed. A good fit to the data with nonzero $\epset$ is achieved by shifting the vacuum oscillation parameters $\theta_{23}$ and $\Delta_{32}^{2}$ (see the definitions below) as shown in Fig~2 of \cite{Friedland:2005vy}. Since the direction of the required shift roughly coincides with the direction of reduced sensitivity at MINOS~\cite{Friedland:2006pi}, even despite recent MINOS data, large values of NSI remain allowed \cite{Maltoni2011}.

Similarly, the bounds from non-oscillation experiments do not exclude values of  $\epset$ hinted at by solar data.  For example, quark-NSI couplings at hadron colliders manifest themselves as anomalous monojet, monophoton~\cite{Friedland:2011za}, and multilepton signals~\cite{Davidson:2011xz,Friedland:2011za}.  Presently, the LHC and the Tevatron are starting to encroach on interesting regions of parameter space for heavy or resonantly produced mediators, but provide essentially no constraint when the mediator mass is $\lesssim 10$ GeV.  (Similar constraints on NSI involving electrons were derived from LEP data using monophoton events~\cite{Berezhiani:2001rs}.) It is worth emphasizing that this ``light mediator'' regime could easily first appear in solar neutrino or long-baseline data. A similar possibility of very light ($\ll$ GeV) mediators in the neutrino sector has been recently invoked in order to explain dark matter direct detection anomalies~\cite{Pospelov:2011ha,Pospelov:2012gm,Harnik:2012ni}, the short-baseline anomalies~\cite{Nelson:2007yq} , and the apparent CPT violation at MINOS~\cite{Engelhardt:2010dx} (which has since been resolved).  

Other constraints on NSI in the contact operator limit, from fixed target experiments and rare decays, are compiled in~\cite{Davidson:2003ha}. For example, for the $\epset$ coupling in question, the CHARM beam-dump experiment~\cite{Dorenbosch:1986tb} constrains $\epset^{q} \lesssim 0.5$, while constraints on hadronic decays of the tau restrict $\epset^{q} \lesssim 1.6$~\cite{Davidson:2003ha}. A precise determination of how such experimental constraints are altered when the mediator of neutrino NSI is allowed to be light is yet to be carried out. 
Moreover, light mediators of NSI can contribute to quark-quark and neutrino-neutrino scattering
Stringent bounds on light vector mediators between quarks exist from low energy neutron-nucleus scattering experiments~\cite{Barger:2010aj}, while supernovae may provide a constraint on neutrino-neutrino scattering. The analysis becomes model-dependent and is beyond the scope of this work.

The further testing of the anomalous matter effect due to NSI should occur with the next generation of solar neutrino experiments, notably SNO+, and with the long-baseline experiments, notably \nova. Both are currently under construction and are expected to start operations next year. After \nova, the LBNE experiment, with a longer 1300 km baseline is expected to come online. The NSI sensitivity of \nova~and LBNE is the subject of the present paper.  As we show below, these experiments may probe NSI couplings at the levels hinted at by the solar data. This is made possible by the recently measured large value of $\theta_{13}$, such that the dominant NSI contribution comes \emph{at linear order}, from interference between standard and non-standard physics. These searches will be especially of great interest should SNO+ confirm the distortion of the $^{8}$B solar neutrino spectrum.

The outline of this paper is as follows. In Sec.~\ref{std} we specify our conventions for the description of standard oscillation physics. In Sec.~\ref{NSIeqs} we introduce the formalism for describing NSI and derive a useful analytic expression for the $\nu_{\mu} \rightarrow \nu_{e}$ conversion probability, highlighting that the dominant effect from NSI comes from interference between standard and non-standard physics. In Sec.~\ref{sec:MINOS} we show that MINOS is at the edge of sensitivity to a portion of the parameter space favored by solar data, though with significant degeneracies of interpretations existing. In Sec.~\ref{sec:nova} we turn to the \nova~ experiment and show that a combined analysis of the neutrino and antineutrino conversion probabilities may reveal a measurable effect of the NSI. We also discuss how the possible presence of NSI degrades \nova's ability to determine the hierarchy and the vacuum phase. 
Lastly,  in Sec.~\ref{sec:LBNE} we look forward to LBNE with its longer baseline as a way to break most of these degeneracies, and over most of the parameter space, deduce the hierarchy and presence or absence of NSI.  Our conclusions are summarized in Sec.~\ref{sec:conc}.

\section{NSI in the $e-\tau$ sector}
\label{sec:NSI}
\subsection{Standard Three-Flavor Vacuum Oscillation Physics}
\label{std}
Measurements from solar, atmospheric, reactor and beam neutrino experiments indicate flavor transformations between the three known flavors of neutrinos.Thus neutrinos are massive and their mass and flavor bases are misaligned.
In the mass basis, the three-flavor oscillation Hamiltonian is parameterized as follows,
\be 
H_{vac}^{mass} = 
 \begin{pmatrix}
  -\Delta_{\odot}-\Delta & 0 & 0 \\
  0 & \Delta_{\odot}-\Delta  & 0 \\
  0 & 0 & \Delta_{\odot}+\Delta 
 \end{pmatrix},
 \label{Hvac2}
\ee
where $\Delta \equiv \Delta m_{32}^{2}/ (4E_{\nu})$, and  $\Delta_{\odot} \equiv \Delta m_{21}^{2}/ (4E_{\nu})$, with $m_{ij}^{2} \equiv m_{i}^{2}-m_{j}^{2}$ and $E_{\nu}$ the neutrino energy. We stress that the restriction to the known three flavors is a simplifying assumption, in keeping with our simplified framework. In general, one may want to consider NSI in the presence of additional sterile neutrino states. 

In writing Eq.~(\ref{Hvac2}) we have used the fact that $E_{\nu}\gg m_{i}$, and dropped both the Majorana phases and an overall constant, since neither contribute to oscillation probabilities, as is well known. Although the magnitude of the atmospheric splitting $|\Delta m_{32}^{2}|$ is well measured by MINOS~\cite{MINOSatm} to be $\left(2.32 ^{+0.12}_{-0.08}\right) \times10^{-3} \rm{eV}^{2}$, and the solar splitting $\Delta m_{21}^{2} =7.1^{+1.2}_{-0.6}\times10^{-5}~\rm{eV}^{2}$ is well measured by KamLAND~\cite{Abe:2008aa}, the sign of the former, which determines the ordering of the three mass states, remains unknown. We follow the prevailing convention in referring to $\Delta >0$ as the normal hierarchy (NH), and $\Delta <0$ as the inverted hierarchy (IH).

Eq.(\ref{Hvac2}) can be rotated to the flavor basis $H_{vac}^{mass} \rightarrow UH_{vac}^{mass}U^\dagger$ via the PMNS matrix. Following the standard notation \cite{PDG2010}, 
\begin{widetext}
\begin{equation*} 
U_{PMNS} = 
 \begin{pmatrix}
  1 & 0 & 0\\
  0 & c_{23}  & s_{23} \\
  0 & -s_{23} & c_{23}
 \end{pmatrix}
 \begin{pmatrix}
  c_{13} & 0 & s_{13}e^{-i\delta} \\
  0 & 1  & 0 \\
  -s_{13}e^{i\delta} & 0 & c_{13}
 \end{pmatrix}
 \begin{pmatrix}
  c_{12}& s_{12} & 0 \\
  -s_{12} & c_{12}  & 0 \\
  0 & 0 & 1
 \end{pmatrix},
 \label{PMNS}
\end{equation*}
\end{widetext}
where $c_{ij} \equiv \cos \theta_{ij}$ and $s_{ij} \equiv \sin \theta_{ij}$. Thus in vacuum neutrino oscillations are characterized by three angles and one $CP$-violating phase $\delta$. We adopt the values $\theta_{23} = 45^{\circ} \pm 7.7^{\circ}$~\cite{MINOSatm}, $\theta_{12} = 34^{\circ} \pm 1.1^{\circ}$~\cite{SNO12}, and the recently updated $\theta_{13} = 8.7^{\circ} \pm 0.5^{\circ} $~\cite{DB2} (see~\cite{DB} for the published results from Daya Bay). The CP-violating phase $\delta$ remains unknown and, together with the type of mass hierarchy, constitutes the main ``known'' target for the next generation of the long-baseline experiments. As we will see below, NSI provide another important target.

\subsection{Non-standard Oscillation Physics}
\label{NSIeqs}

When neutrinos propagate in matter, their coherent forward scattering on the medium (refraction) must be taken into account~\cite{Wolfenstein:1977ue}.  The presence of NSI between neutrinos and quarks (and/or electrons) in the medium modifies this matter potential, thereby altering the neutrino oscillation probability in long-baseline experiments.  Consistent with our framework,  we turn on only non-standard $e-\tau$ interactions,
\be 
H_{mat}^{flav} = 
\sqrt{2} G_{F}n_{e} \begin{pmatrix}
 1 & 0 & |\varepsilon_{e\tau}|~e^{-i\delta_{\nu}} \\
  0 & 0  & 0 \\
  |\varepsilon_{e\tau}|~e^{i\delta_{\nu}} & 0 & 0
  \end{pmatrix},
  \label{Hmatt}
\ee
where $G_{F}$ is the Fermi constant, and $n_{e}$ is the number density of electrons. 
For the experiments considered here, the neutrino beams stay inside the continental crust, for which we take $n_{e} = 1.4~ \rm{mol}/\rm{cm}^{3}$. The ``11'' entry of Eq.~(\ref{Hmatt}) is the SM piece, while $\varepsilon_{e\tau}$ parameterizes the strength of NSI in the electron-tau sector. As already mentioned, the NSI piece can come from electrons, up or down quarks, or a combination thereof. We therefore define $\epset \equiv \sum_{f=e,u,d} \varepsilon_{e\tau}^{f}\langle n_{f} \rangle/n_{e}$, such that the NSI couplings in Eq.~(\ref{Hmatt}) are normalized per electron.  As discussed in the introduction, the NSI Lagrangian term we have in mind is a effective operator of the form, $\mathcal{L} \supset \
-2 \sqrt{2}~ \varepsilon_{e\tau}^{f} G_{F}~ \left(\overline{f} \gamma_{\mu}  f \overline{\nu}_{e} \gamma^{\mu} \nu_{\tau}\right) + h.c.$ This is a useful parameterization for our purposes since many possible UV completions lead to such an operator at low energies. We leave the model-building and non-oscillation phenomenology of such UV completions to a future study. 

The rationale for choosing to consider only $\epset$ is as follows: (i) The interactions involving the muon neutrino are more strongly constrained than those of either $\nu_{e}$ or $\nu_{\tau}$. (ii) Small flavor-diagonal (FD) NSI for $\nu_{e}$ and/or $\nu_{\tau}$ would lead to a small modification of the diagonal matter potential, as if the neutrinos travel through matter of slightly different densities, while small flavor-changing (FC) NSI can lead to qualitatively new effects, as will be seen later. (iii) Large FD NSI could be allowed in combination with suitably chosen FC NSI \cite{Alex2,Friedland:2005vy}, but as mentioned above, in this paper, for simplicity we do not consider cancellation between several NSI parameters.

Observe that the NSI piece in Eq.~(\ref{Hmatt}) is in general complex, with a new irremovable $CP$ violating phase $\delta_{\nu} \equiv \arg \left( \varepsilon_{e\tau}\right)$~\cite{GonzalezGarcia:2001mp,Campanelli:2002cc,Friedland:2004pp}.  We shall see that that this phase interferes with the $\delta$ phase of the ``atmospheric'' oscillation term, producing important effects and fundamental degeneracies.

The measured value of $\theta_{13}$ and the assumed smallness $\epset$ leads us to anticipate the concomitant smallness of the probability $P\left( \nu_{\mu} \rightarrow \nu_{e}\right)$ which we will verify \emph{a posteori}. With the $\nu_{e}$ appearance probability assumed small we can treat the problem in perturbation theory, resulting in~\cite{Friedland:2006pi}:
\begin{widetext}
\bea P\left( \nu_{\mu} \rightarrow \nu_{e}\right) &\approx &\left| G_{1} \sin \theta_{23} \frac{e^{i\Delta_{1}L}-1}{\Delta_{1}} - G_{2} \cos \theta_{23} \frac{e^{i\Delta_{2}L}-1}{\Delta_{2}}\right |^{2}, ~~ \label{prob} \\
G_{1} &=& \sqrt{2} G_{F}n_{e} |\varepsilon_{e\tau}| e^{i\delta_{\nu}} \cos \theta_{23} + \Delta \sin 2\theta_{13}e^{i\delta}, \\
G_{2} &=& \sqrt{2} G_{F}n_{e} |\varepsilon_{e\tau}| e^{i\delta_{\nu}} \sin \theta_{23} - \Delta_{\odot} \sin 2\theta_{12},
\eea
\end{widetext}
with $\Delta_{1,2} \simeq \Delta \pm |\Delta| -\sqrt{2} G_{F}n_{e}$ (neglecting the solar splitting $\Delta_{\odot}$, see below), and $L$ the distance of propagation. Compared to~\cite{Friedland:2006pi} we have taken only $|\varepsilon_{e\tau}|$ nonzero and restored explicitly the phases of both the vacuum and the NSI pieces.
For typical energies $E_{\nu} = 2$ GeV, $\theta_{23} = \pi/4$, and $\theta_{13}=8.7^{\circ}$, the relevant parameters in the problem are 
\bea
\Delta \sin 2 \theta_{13} &=&0.87\times10^{-13}~ \rm{eV}, \label{atmterm}\\
\sqrt{2} G_{F}n_{e} \cos \theta_{23} &=& 0.76 \times 10^{-13}~\mbox{eV},\label{NSIterm} \\
\Delta_{\odot} \sin 2\theta_{12} &=& 0.09 \times 10^{-13}~\mbox{eV} \label{solterm}.
\eea
%


\begin{figure}[t]
  \includegraphics[angle=0,width= \columnwidth]{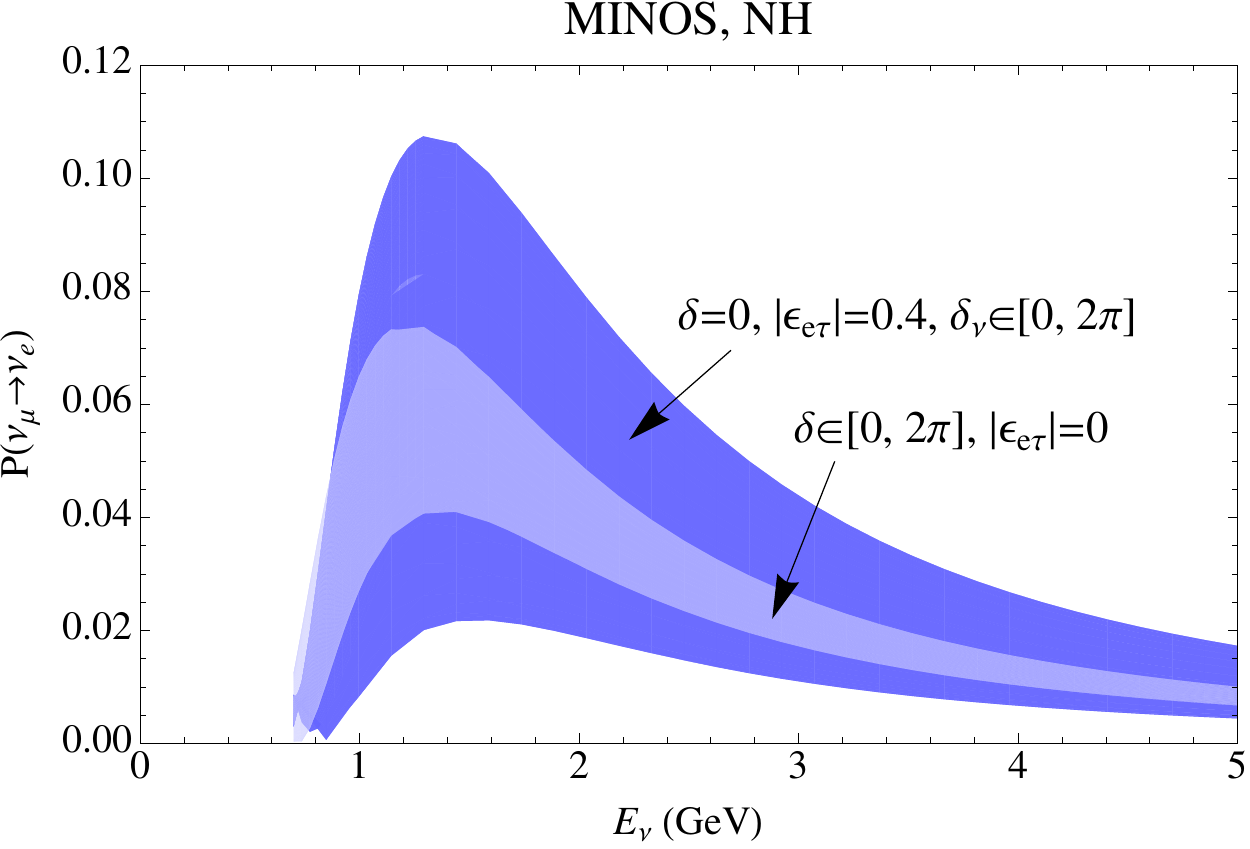} \\
  \includegraphics[angle=0,width= \columnwidth]{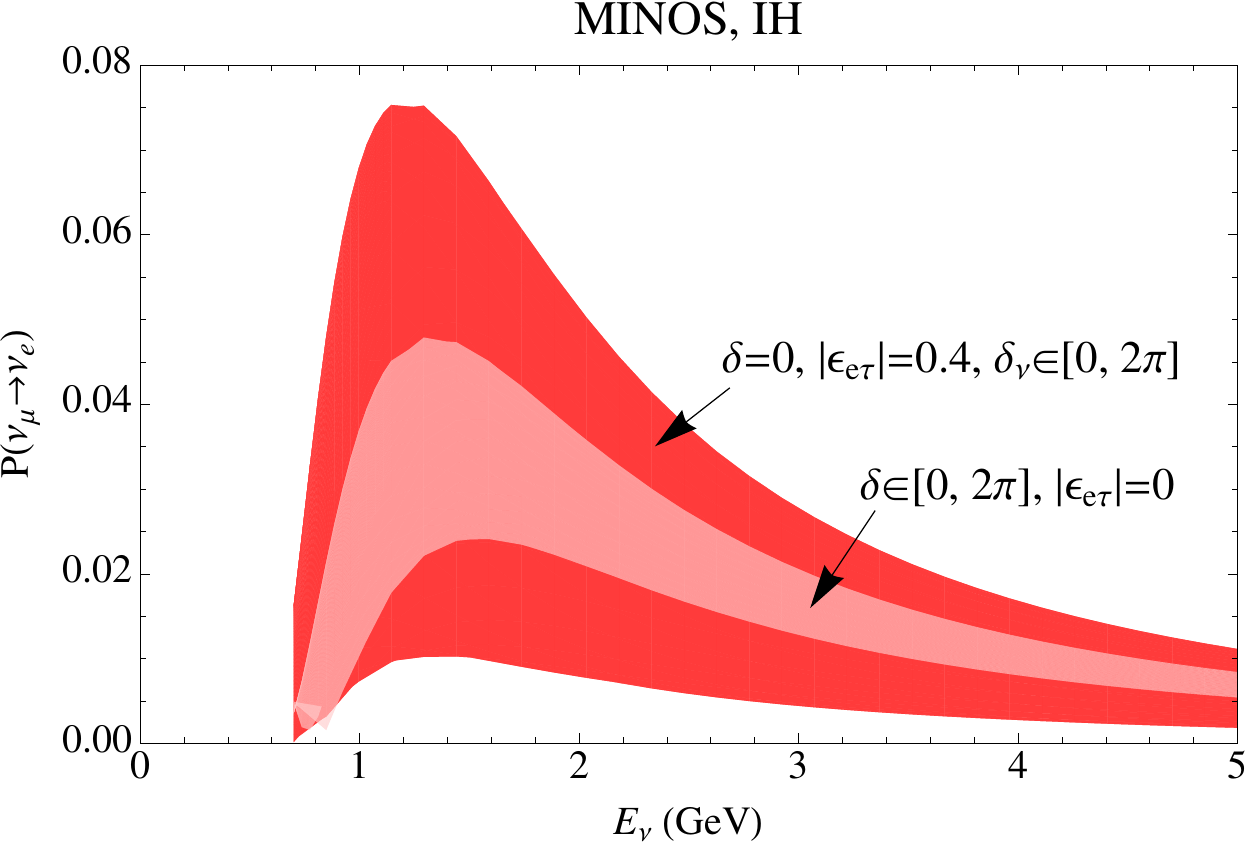}
  \caption{Here we examine MINOS sensitivity to NSI in the normal (upper panel) and inverted hierarchy (lower).  The lighter region comes from varying the vacuum phase $\delta$ with SM physics only, while the darker regions come from including NSI with $|\epset| = 0.4$ and varying the NSI phase with the vacuum phase set to zero.}
  \label{fig:MINOS}
\end{figure}

The physics behind the general form of Eq.~(\ref{prob}) can be understood as follows. The $\nu_{\mu} \rightarrow \nu_{e}$ conversion amplitude receives contributions from two frequencies, related to the ``atmospheric'' ($\Delta_{1}\simeq2\Delta$) and the smaller ``matter'' ($\Delta_{2}\simeq-\sqrt{2}G_{F}n_{e}$) splittings (the ``solar'' $\Delta_{\odot}$ is smaller still and is for simplicity neglected). In the standard case (no NSI), the term $\Delta\sin2\theta_{13}e^{i\delta}$ drives the conversion $\nu_{\mu} \rightarrow \nu_{e}$ with the atmospheric oscillation frequency, as captured by the $G_{1}$ term in Eq.~(\ref{prob}). The smaller off-diagonal ``solar'' term $\Delta_{\odot}\sin2\theta_{12}$, captured by the $G_{2}$ term, also drives the transition, but with a smaller frequency, $\Delta_{2}$. 

The standard $CP$ violation search is based on the interference of the terms in Eqs.~(\ref{atmterm}) and (\ref{solterm}). The magnitude of interference is dictated by the phase $\delta$, which is responsible for $CP$ violation, and by the oscillation phases, $\arg\left (e^{i\Delta_{1,2}L}-1\right)$, in the two channels. Furthermore, since the solar term (\ref{solterm}) is an order of magnitude smaller than the atmospheric one, $CP$ violation appears as a subleading effect, modifying the leading probability due to Eq.~(\ref{atmterm}) by at most $\sim20\%$ (when the interference is maximally constructive or destructive). Its observation thus requires sufficient experimental precision. 

The presence of nonzero $\epset$ NSI modifies the amplitudes of both channels. 
Physically, ordinary oscillations generate $\nu_{\tau}$, which $\epset$ then converts into $\nu_{e}$. 
With $|\varepsilon_{e\tau}|$ of order 0.2-0.4 and $E_{\nu}\sim2$ GeV, the hierarchy of terms in Eqs.~(\ref{atmterm},\ref{NSIterm},\ref{solterm}) becomes: $atm > NSI> sol$. Thus, the expected NSI effect is still subleading, but in general \emph{larger} than the standard signal of $CP$ violation.
The observable effect of NSI then depends at leading order on the \emph{relative phase} $\delta_{\nu} - \delta$. 
As an illustration, when this relative phase is zero and $|\epset|=0.2$, one expects a $\sim30\%$ enhancement on top of the leading atmospheric probability.  


We finish this section with two important corollaries to the above discussion. First, since ordinary oscillations form the necessary first stage of the conversion, at high neutrino energy $E_{\nu}$, the $\nu_{e}\rightarrow\nu_{\mu}$ conversion probability goes to zero even in the presence of nonzero $\epset$. Thus, while naively one might expect nonstandard matter effects to be cleanly manifested at high energies\footnote{Which is indeed true for, {\it e.g.}, $\varepsilon_{\mu\tau}$. In that case, one expects $\nu_{\mu}\rightarrow\nu_{\tau}$ conversion at high energy.}, this is not so for $\epset$. The best energy to probe $\epset$ in $\nu_{\mu} \rightarrow \nu_{e}$ conversion is at the appearance maximum.
Thus, \nova~(and and its proposed successor, LBNE) are suitable experiments to look for this type of new physics.

The second observation is that the recent measurement of large $\theta_{13}$ is crucial in giving MINOS, NO$\nu$A, and LBNE sensitivity to NSI, since the NSI-driven conversion interferes with the ``standard'' amplitude driven by $\theta_{13}$.  As an illustration, consider the fact that with $\theta_{13} = 0$ and $|\varepsilon_{e\tau}| =0.2$ this probability at NO$\nu$A is below $0.005$, even with constructive NSI-solar interference. This signal is certainly below the sensitivity reach of the next generation of long-baseline experiments. To get an observable signal at MINOS, $P\left( \nu_{\mu} \rightarrow \nu_{e}\right)\sim 0.05$, with $\theta_{13} = 0$ requires large NSI, $|\varepsilon_{e\tau}| \approx 0.9$~\cite{Friedland:2006pi}.  Since in the past year the value of $\theta_{13}$ was measured to be sufficiently large, it is time to revisit the sensitivity of MINOS to NSI.


\section{MINOS}
\label{sec:MINOS}
Of the long-baseline oscillation experiments that already have data, MINOS provides the best sensitivity to NSI. This is due to their relatively long baseline and the resolution of their $P(\nu_{\mu} \rightarrow \nu_{e})$ measurements. In fact, as we show below, the $\nu_{e}$ appearance search by MINOS~\cite{Adamson:2011qu} has already started approaching the region of the parameter space favored by solar data.  

We begin by asking what the NSI sensitivity of MINOS is in the case of the normal hierarchy, $\Delta > 0$.  To this end, Eq.~(\ref{prob}) is plotted for the MINOS baseline in the top panel of Fig.~\ref{fig:MINOS}, both with NSI of magnitude $|\epset| =0.4$ (darker region) and without it (lighter region).  The standard physics region is swept out by the variation of the vacuum phase $\delta$, while the NSI region has vanishing vacuum phase but varying the $\delta_{\nu}$ phase. All other oscillation parameters are set to their measured central values (see Sec.~\ref{std}). From Fig.~\ref{fig:MINOS} we see that in the presence of the NSI the two hierarchies overlap considerably. This is the first example, of many to be encountered, illustrating the ability of NSI to easily confuse the interpretation of experimental data when viewed in terms of standard oscillation physics. 

Furthermore notice that when NSI interferes constructively with the SM contribution, the $\nu_{\mu} \rightarrow \nu_{e}$ conversion probability can be as large as $\sim0.11$. At face value, this is in conflict with the $90 \%$ CL bound from MINOS~\cite{Adamson:2011qu}.  However, the sensitivity of MINOS diminishes appreciably after marginalizing over the uncertainty in known oscillation parameters. Indeed, recall that Fig.~\ref{fig:MINOS} does not include the allowed variation in the standard oscillation parameters, $\theta_{13}$ and $\theta_{23}$. With the now small error bars on $\theta_{13}$ from Daya Bay, tuning $\theta_{13}$ down does not change the conversion probability appreciably.  Suppose, however, that the actual $\theta_{23}$ value in nature is $1 \sigma$ away from its central value, $\theta_{23} \approx 37^{\circ}$. This renders the largest NSI probability $P\left(\nu_{\mu} \rightarrow \nu_{e}\right) \approx 0.083$. This conversion probability is only $1\sigma$ higher than the MINOS best-fit point in the NH. A similar exercise in the case of the IH does not yield any exclusion of the solar-preferred values of NSI.



Notice, additionally, that the $\nu_{e}$ appearance probability in a given hierarchy can mimic that of the other hierarchy.  
Comparing the two panels of Fig.~\ref{fig:MINOS} one observes that the presence of NSI significantly exacerbates the difficulty in determining the hierarchy.  Folding in information regarding the location of the peak would certainly increase the sensitivity of MINOS to the sign of the hierarchy, since the NH peaks at somewhat lower energies.  Given their substantial background however, MINOS appears unlikely to have sufficient sensitivity to this signature of the hierarchy. 

{\it A priori}, MINOS could have observed indications of anomalous conversion due to NSI $|\epset|\sim0.4$, had the phases been such that the interference is constructive and the true value of $\theta_{23}$ were on the high end of the allowed range. This shows that MINOS is approaching an interesting range of sensitivities.
With MINOS already having skimmed off a small portion of the NSI parameter space thanks to large $\theta_{13}$, it is natural to ask how this progress will be improved upon. The next relevant experiment is \nova, which is designed to improve upon the MINOS $\nu_{\mu}\rightarrow\nu_{e}$ results by about an order of magnitude. Thus, it is timely to examine the effect of the NSI at ~\nova.

\begin{figure}[h]
\includegraphics[width=0.8\columnwidth]{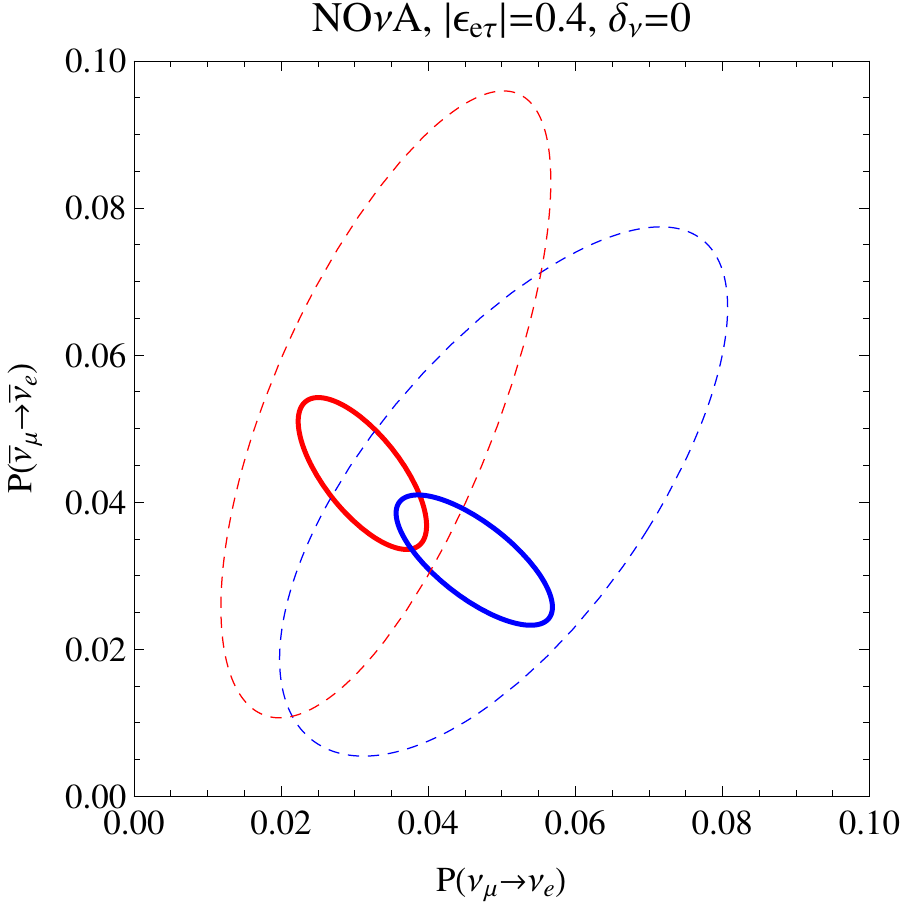}
  \caption{Here we have fixed the neutrino energy to $2~$GeV, and plotted the ensuing values of $P(\nu_{\mu}\rightarrow\nu_{e})$ and $P(\overline{\nu}_{\mu}\rightarrow \overline{\nu}_{e})$ for NO$\nu$A. The solid ellipses refer to SM only interactions, for the normal (blue) and inverted (red) hierarchy, with the vacuum phase $\delta$ varying continuously along each of the ellipses. For the dashed ellipses we have used the NSI values $|\varepsilon_{e\tau}| = 0.4$ and $\delta_\nu=0$, corresponding to the solar neutrino fit in Fig.~\protect{\ref{fig:SNO}} (solid curve there). }
  \label{fig:NOvAbiprob1}
\end{figure}


\section{NO$\nu$A}
\label{sec:nova}
With a baseline of 810 km, a narrow-band 2 GeV beam around the expected $\nu_{\mu}\rightarrow\nu_{e}$ conversion maximum and enhanced detector sensitivity, the NO$\nu$A experiment~\cite{novaproposal} will offer the next opportunity to measure or constrain NSI.  Supposing NO$\nu$A runs for 3 years in each mode, we ask what additional information can be gleaned from combining data from both neutrinos in $P(\nu_{\mu}\rightarrow\nu_{e}) \equiv P$ and antineutrinos in $P(\overline{\nu}_{\mu}\rightarrow\overline{\nu}_{e}) \equiv \overline{P}$.  

A particularly simple way of depicting this is through the use of a ``bi-probability plot'' \cite{Minakata:2001qm}. This method has been widely used by the \nova~collaboration for presenting their standard physics expectations. With only SM interactions, a bi-probability curve is constructed by fixing the neutrino energy $E_{\nu}$ and the mass hierarchy and varying the vacuum phase $\delta$.  This ``theory only'' prediction can be used to understand the required experimental sensitivity to the phase $\delta$ and the sign of $\Delta$.  Notice that although a bi-probability plot only includes information from a single energy bin, for NO$\nu$A this is not a significant limitation since it is by design a narrow-beam experiment with $E_{\nu}\sim 2$ GeV ({\it cf.} Fig.~4.5 in \cite{novaproposal}).  

The predicted bi-probability curves for the NO$\nu$A baseline are depicted in Fig.~\ref{fig:NOvAbiprob1}. The solid curves assume standard oscillations physics only, with the blue ellipse referring to the NH and the red ellipse to the IH. The perimeter of each curve is traced from the variation of the vacuum phase, $0 \le \delta \le 2\pi$. The points closest to the center of each ellipse correspond to no CP violation ($\delta=0$ or $\pi$), while those farthest from the center correspond to maximal violation ($\delta=\pi/2$ or $3\pi/2$), {\it cf.} Fig.~3.3 in \cite{novaproposal}. Notice that as $\delta$ runs through its range, the probabilities are modulated by at most $\sim0.01$, or about $20\%$. This is indeed expected from the interference of the solar term with the leading atmospheric term, as discussed in Sect.~\ref{NSIeqs}.

This ``theory only'' plot could in fact serve as a guide for the experimental sensitivity, if combined with the expected resolution information. The collaboration projects its $2\sigma$ resolution to be roughly a circle in the bi-probability plane, with the diameter of $\sim0.15$ \cite{novaneutrino2012}.
 For example, by measuring $P \sim 0.05$ and $\overline{P}\sim 0.03$, \nova~could claim to have evidence that the hierarchy is normal at  almost $3\sigma$ confidence level (C.L.). Moreover, at $2\sigma$ C.L., the phase $\delta$ would be determined to be $3\pi/2\pm\pi/2$.  As a second example, if \nova~were to measure $P \sim \overline{P} \sim 0.04$, the hierarchy could not be established. It would be known, however, that the $\delta$ phase is either in the interval  $\pi/2\pm\pi/2$ with NH, or in the interval $3\pi/2\pm\pi/2$ with NH. In other words, the existence of CP violation would be suggested at the $2\sigma$ C.L. 

Crucially, all these determinations would apply only if it were somehow known that NSI was ruled out.
When we allow for nonzero NSI in the $e-\tau$ sector, the situation changes considerably, and the regions corresponding to the two hierarchies expand significantly. To illustrate this, we first consider the NSI scenario that was used in the fit of the solar data in Fig.~\ref{fig:SNO}. Recall that the thick curve in that figure was constructed for $|\epset|=0.4$ and the phase $\delta_{\nu}=0$. In this case, at \nova~the variation of the vacuum phase traces out the dashed ellipses in Fig.~\ref{fig:NOvAbiprob1}, analogously to the SM ellipses discussed above.  When NSI-SM interference is constructive the probabilities in both neutrino and antineutrino modes can be substantially larger, approaching $0.08-0.10$. If nature prefers such fortuitous values of NSI, NO$\nu$A will see dramatic deviations from the SM expectations. We see that NO$\nu$A is in an excellent position to probe NSI couplings at a level suggested by the solar data.

Notice that the size of the NSI effect in Fig.~\ref{fig:NOvAbiprob1} can be once again roughly understood from the discussion in Sect.~\ref{NSIeqs}: since $\sqrt{2} G_{F}n_{e} \cos \theta_{23}\epset$ is $\sim3$ times larger than  $\Delta_{\odot} \sin 2\theta_{12}$, the dashed ellipses are roughly a factor of three longer.

\begin{figure}[t]
\includegraphics[width=0.8\columnwidth]{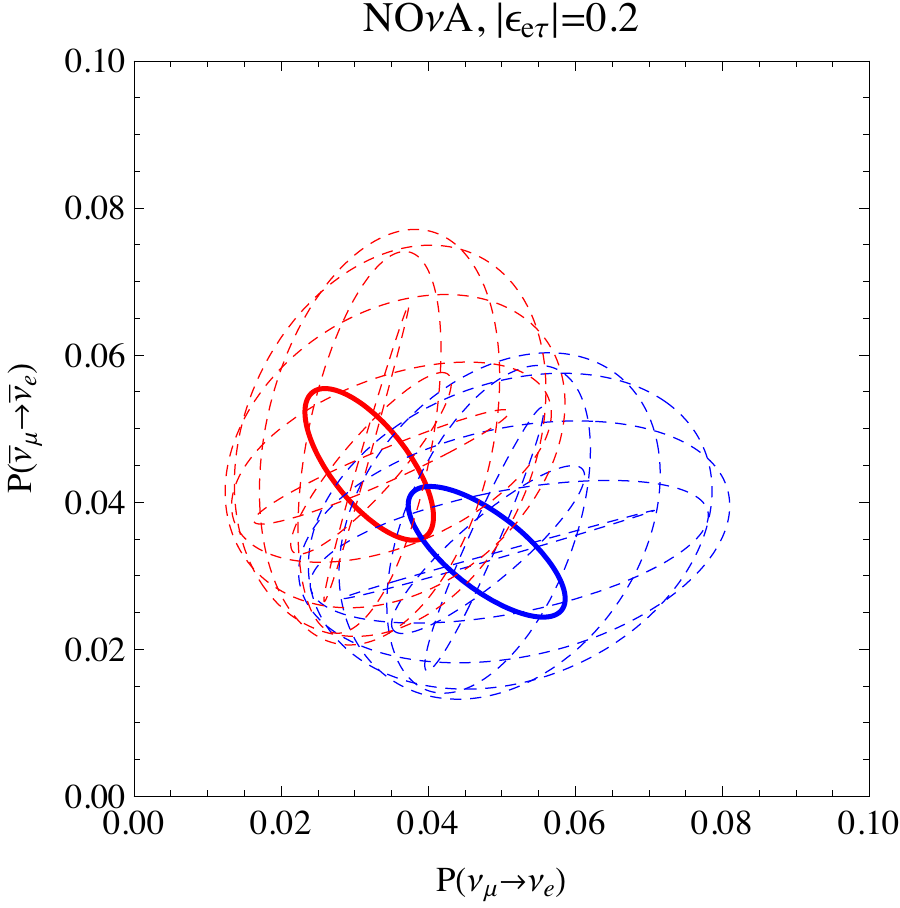}
\includegraphics[width=0.8\columnwidth]{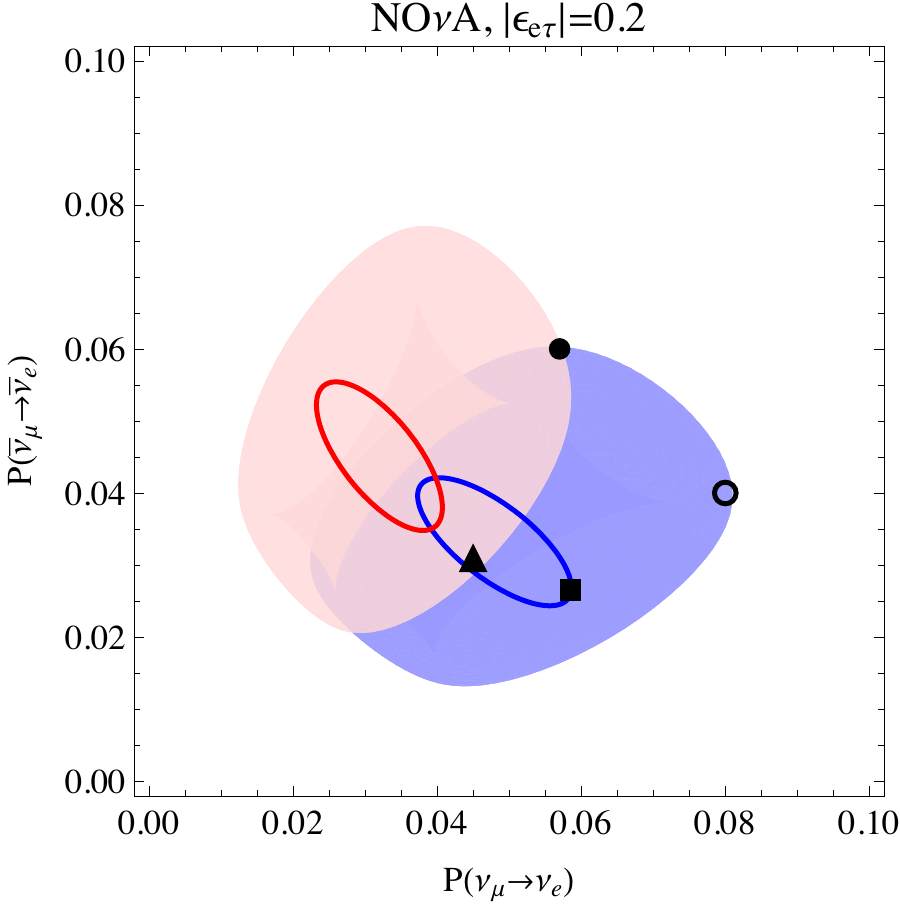}
  \caption{ Similar to  Fig.~\ref{fig:NOvAbiprob1}, although here we fix $|\epset| =0.2$. \emph{Top:} Scanning over many values of the NSI phase $\delta_{\nu}$ results in significant degeneracy, with many distinct curves intersecting at a given point. \emph{Bottom:} The result of allowing $\delta_{\nu}$ to float arbitrarily yields the shaded blobs. See text for a discussion of the four qualitatively distinct points.}
  \label{fig:NOvAbiprob2}
\end{figure}

Alternatively, one may be interested in the general question of \nova's sensitivity to $\epset$, without reference to the solar data. To address this issue, we vary the $\delta_{\nu}$ phase in its entire range. We also restrict $|\epset|$ to a smaller value, 0.2, which is justified given the level of sensitivity seen in Fig.~\ref{fig:NOvAbiprob1}. With this new fiducial value, we can repeat the same exercise as before but now with several different values of the $\nu$-phase. Doing so in Fig.~\ref{fig:NOvAbiprob2}, we find the regions shown in Fig.~\ref{fig:NOvAbiprob2} (\emph{top}).


Doing this for all possible $\nu$-phases from $0$ to $2\pi$ traces out the large blobs for the two hierarchies, see the bottom panel of Fig.~\ref{fig:NOvAbiprob2}.  Note that for any given point in the NSI regions, many ellipses with distinct vacuum and $\nu$-phases intersect. 
This means considerable degeneracy of parameters corresponding to distinct values of both the $\nu$-phase and the vacuum phase $\delta$ intersecting at any given point. This is especially true in the central regions where one can see what in mathematical jargon is known as a double ``swallowtail catastrophe,'' as the parameter region folds on itself. Based on this observation alone, NO$\nu$A will have difficulty in knowing what combination of phases they are truly measuring. 

Let us now consider four qualitative possibilities for the outcome of NO$\nu$A as illustrated in Fig.~\ref{fig:NOvAbiprob2} (\emph{bottom}): 
\begin{enumerate}
\item[(1)] \emph{Clear NSI and hierarchy determination.} For example, the point $(P,\overline{P}) = (0.08,0.04)$ ($\circ$ in Fig.~\ref{fig:NOvAbiprob2}) is widely discrepant from any SM-only explanation. Moreover, such a set of probabilities would not be likely to have come from SM or NSI cases in the inverted hierarchy. In this case, one could fairly say that the normal hierarchy and nonzero NSI are strongly favored. This represents an example of the best case scenario for NO$\nu$A being sensitive to this new physics. 
\item[(2)] \emph{Clear NSI determination only.} Here the point $(P,\overline{P}) = (0.055,0.06)$ ($\bullet$ in Fig.~\ref{fig:NOvAbiprob2}), would indicate a strong preference for the existence of NSI. Of course with this measurement alone however, no confident statement about the sign of the mass hierarchy could be made.
\item[(3)] \emph{Hierachy determination only.} An exemplary point of this possibility is offered by $(P,\overline{P}) = (0.06,0.03)$ ($\blacksquare$ in Fig.~\ref{fig:NOvAbiprob2}). Here the normal hierarchy would be mildly preferred over any explanation based on the inverted hierarchy. However, this point is degenerate with SM and NSI explanations. 
\item[(4)] \emph{No NSI or hierarchy determination.} The point $(P,\overline{P}) = (0.04,0.03)$ ($\blacktriangle$ in Fig~\ref{fig:NOvAbiprob2}) is an example of one of the worst cases for NO$\nu$A to have a clear signal of any as of yet unknown parameters. At such a point, one cannot rule out the existence of NSI or establish the sign of the hierarchy. 
\end{enumerate}
%


\begin{figure}[t]
  \includegraphics[width=\columnwidth]{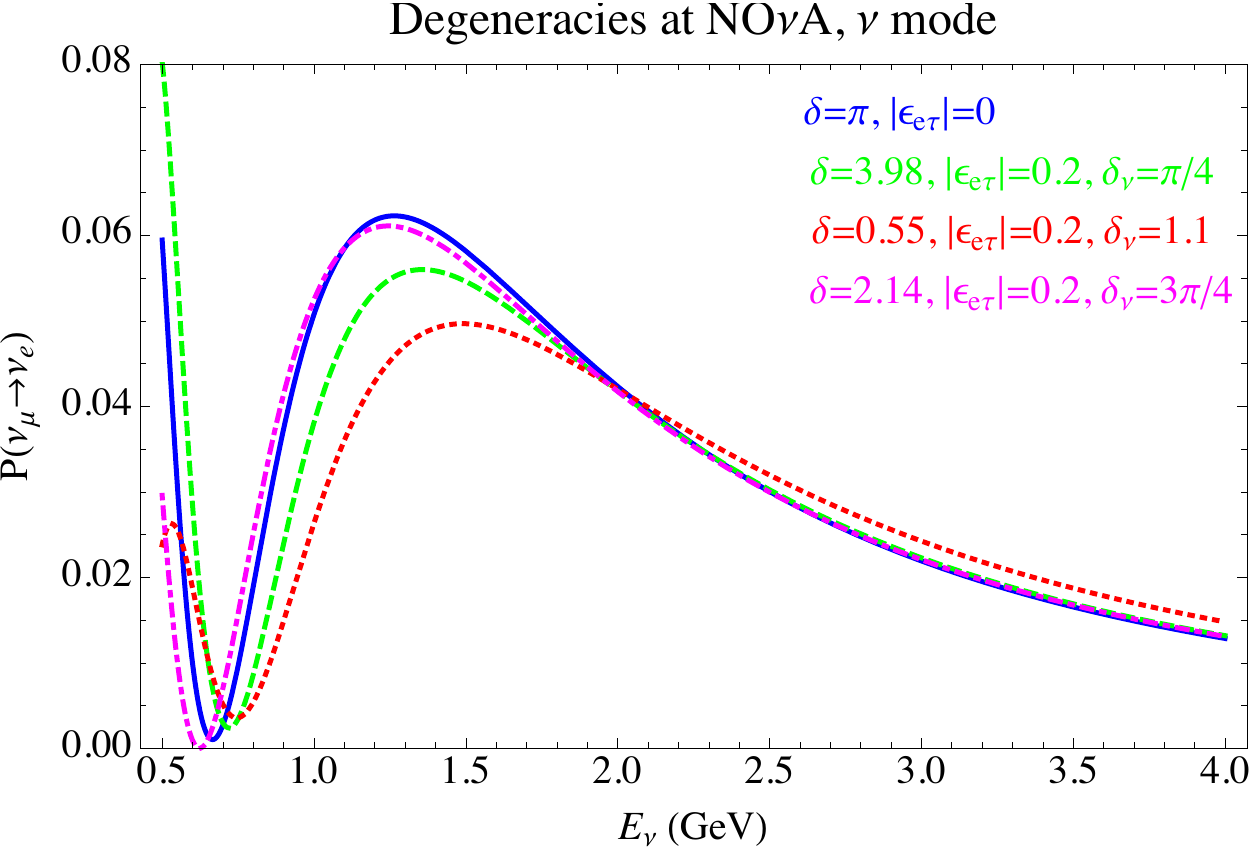} 
   \includegraphics[width=\columnwidth]{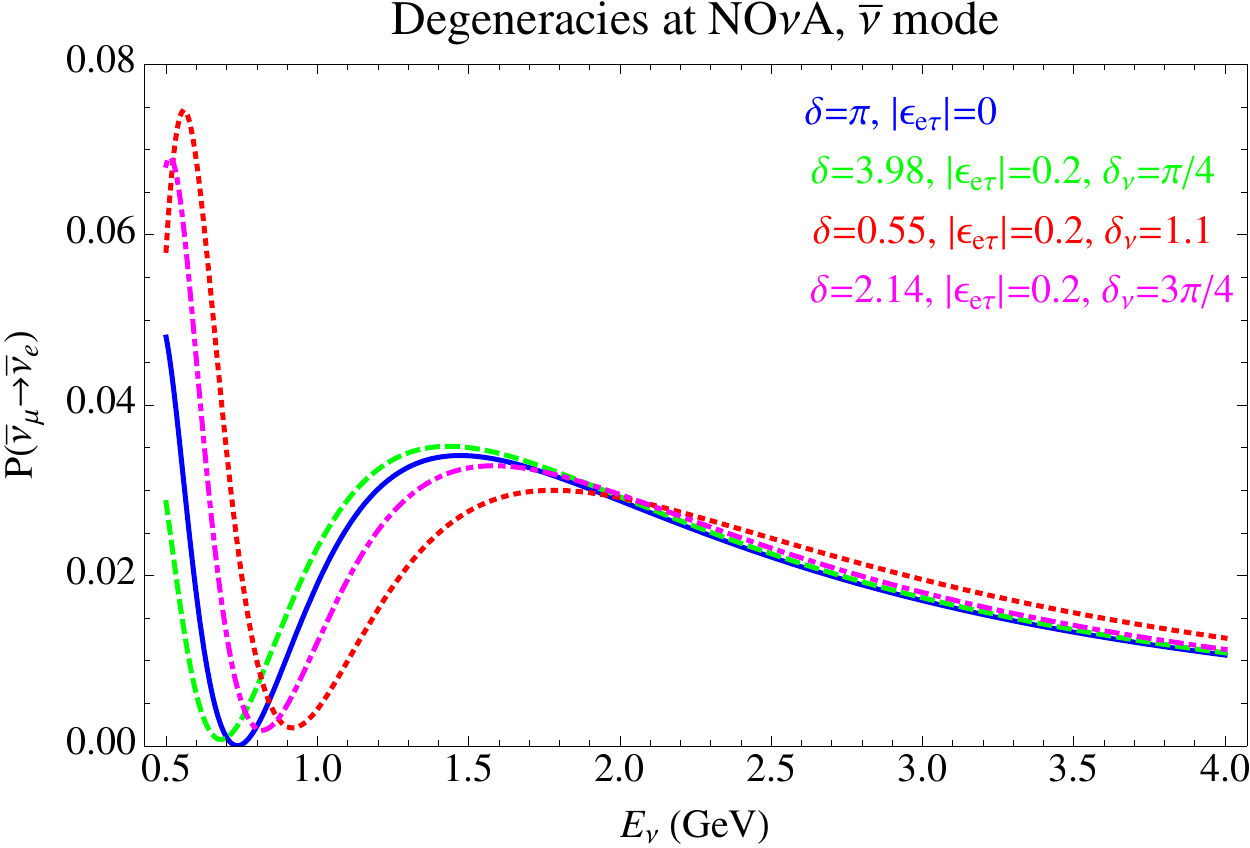} 
  \caption{ An illustration of some of the NO$\nu$A degeneracies at 2 GeV in $\nu$ mode (\emph{top}) and $\overline{\nu}$ mode (\emph{bottom}). All curves assume the normal hierarchy. The blue (solid) curve is without NSI and setting the vacuum phase to $\delta = \pi$. The remaining curves all have NSI of the same magnitude $|\epset| = 0.2$, but take $(\delta, \delta_{\nu}) = (3.98,\pi/4), (0.55, 1.1), (2.14,3\pi/4)$, in the green (dashed), red (dotted), and magenta (dash-dotted) curves respectively. Note that the choice of phases produce degenerate results at 2 GeV in both neutrinos and antineutrinos. }
  \label{fig:NOvAdeg}
\end{figure}

We further illustrate the degeneracy of the last point $\blacktriangle$ in Fig.~\ref{fig:NOvAdeg}, where we plot the conversion probability as a function of energy. One of the curves has only standard physics and $\delta = \pi$, while the other three curves have NSI with different choices of the vacuum and $\nu$-phase. Both neutrino and antineutrino curves intersect at $E_{\nu}=2$ GeV and with a narrow band setup of \nova~give the same event rates. Importantly, the degeneracy is not absolute and can be broken by measurements at different energies and/or baselines. We will return to this in the next section.


\begin{figure}[t]
  \includegraphics[width= \columnwidth]{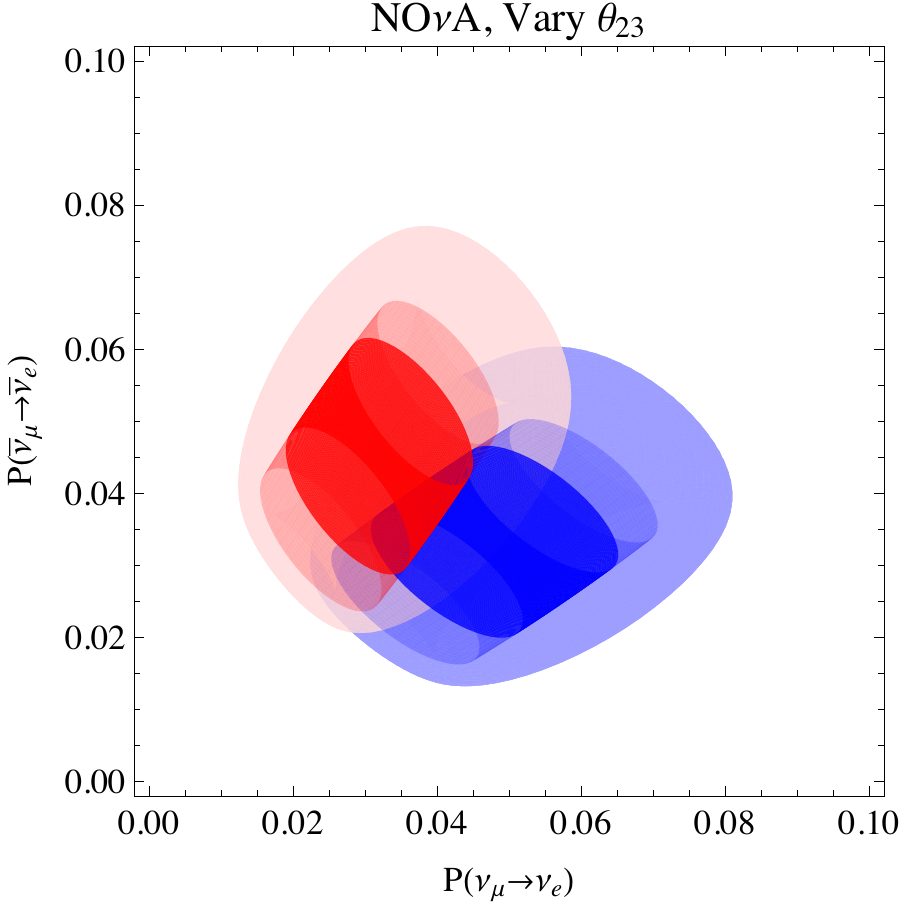}
  \caption{ Here we have fixed the neutrino energy to $2~$GeV, and plotted the ensuing values of $P(\nu_{\mu}\rightarrow\nu_{e})$ and $P(\overline{\nu}_{\mu}\rightarrow \overline{\nu}_{e})$ for NO$\nu$A. The outer (inner) cylinder regions refer to SM only interactions, for the normal (blue) and inverted (red) hierarchy, the angle $\theta_{23}$ varying within its presently (large, light cylinders) and future (small, dark cylinders) allowed 90$\%$ CL region centered on $\pi/4$. The larger shaded regions come from fixing $|\varepsilon_{e\tau}| = 0.2$ and varying both the vacuum and the matter phases.}
  \label{fig:NOvA23}
\end{figure}

Before concluding our discussion of \nova, it is worth mentioning yet another type of degeneracy, which exists between the standard oscillation parameter $\theta_{23}$ and  NSI. 
The underlying physics behind this degeneracy is evident from the analytical form of the probability in Eq.~(\ref{prob}): the change of $G_{1}$ due to NSI can be partially undone\footnote{The degeneracy is clearly partial, even for unconstrained $\theta_{23}$, since $\sin2\theta_{23}$ is real, while $G_{1}$ is complex.} by appropriately modifying the factor of $\sin2\theta_{23}$. 
We have already seen an example of this degeneracy in the case of MINOS, where the effect of NSI could be partially undone by adjusting the value of $\theta_{23}$. 
Let us now describe, quantitatively, this degeneracy at \nova.

Once again, we turn to the bi-probability plane. The effect of varying $\theta_{23}$ in this plane -- assuming standard physics only -- is to shift the solid ellipses in Fig.~\ref{fig:NOvAbiprob1} toward and away from the origin ({\it cf.} \cite{novaneutrino2012}). In other words, varying $\theta_{23}$ turns the ellipses into cylinders. We illustrate this in Fig.~\ref{fig:NOvA23}, which generalizes Fig.~\ref{fig:NOvAbiprob2} to the case of uncertain $\theta_{23}$. The filled background regions are once again obtained by varying $\delta$ and $\delta_{\nu}$, assuming fixed $|\epset| =0.2$ and $\theta_{23}=\pi/4$. The lightly shaded foreground cylinders are the result of setting $|\epset|$ to zero and varying $\delta$ over its full range and $\theta_{23}$ over the range allowed by the Super-Kamiokande atmospheric data\footnote{The values of $\sin^{2} 2\theta_{23}$ inferred from the Super-Kamiokande data actually change for large NSI, as shown in \cite{Alex2,Friedland:2005vy}. We neglect this change here, since we work in the assumption of small NSI (no cancellations between different $\varepsilon$'s).} $\sin^{2} 2\theta_{23} > 0.93$. We see that the degeneracy between NSI and $\theta_{23}$, while only partial, reduce the possible statistical significance of the NSI signal. 

It then becomes crucial for the NSI search to have a better determination of $\theta_{23}$. Fortunately,~\nova~itself will be able to improve the error on $\theta_{23}$ by about a factor of two, by using data in the $P(\nu_{\mu}\rightarrow\nu_{\mu})$ channel \cite{novaneutrino2012}. This results in the smaller darker cylinders also seen in Fig.~\ref{fig:NOvA23}, which correspond to $\sin^{2} 2\theta_{23} > 0.98$. Thus, the improved measurement of $\theta_{23}$ by \nova~ translates into improved NSI sensivity.

Further improvements could be achieved in two ways: by tightening the bounds on $\theta_{23}$ from other measurements and by going to a longer baseline. The latter possibility is described in the next section. As for the former avenue, an important contribution could be made by the Deep Core component of IceCube. The authors of~\cite{Giordano:2010pr} estimate that with 10 years of data, Deep Core will be able to measure $\theta_{23}$ to within $\pm~ 2^{0}$ from maximal mixing. This would be an enormous help in conjunction with the data from NO$\nu$A in measuring or constraining NSI and the sign of the hierarchy. 

The flip side of the $\theta_{23}-$NSI degeneracy is that under the assumption of NSI, \nova~loses any power of ``octant'' discrimination ({\it i.e.}, discrimination between the \emph{light} side, $\theta_{23}<\pi/4$, and the \emph{dark} side, $\theta_{23}>\pi/4$, in the terminology of \cite{darkside}). For example, a point like $(P,\overline{P}) = (0.06, 0.04)$, which under the standard assumptions could be taken as an indication that $\theta_{23}$ is in the dark side, could instead be coming from NSI and $\theta_{23}$ on the light side. 

Of course, uncertainties on the precise value of $\theta_{13}$ remain as well, though these do not produce effects nearly as big as the uncertainties on $\theta_{23}$. This is due to the fact that $\theta_{13}$ is known to within 6$\%$~\cite{DB2} already, whereas $\theta_{23}$ is known only within 17$\%$. Importantly, since the Daya Bay baseline is short, its measurement of $\theta_{13}$ is not impacted by the existence of NSI.


\begin{figure}[t]
    \includegraphics[width=0.45\textwidth]{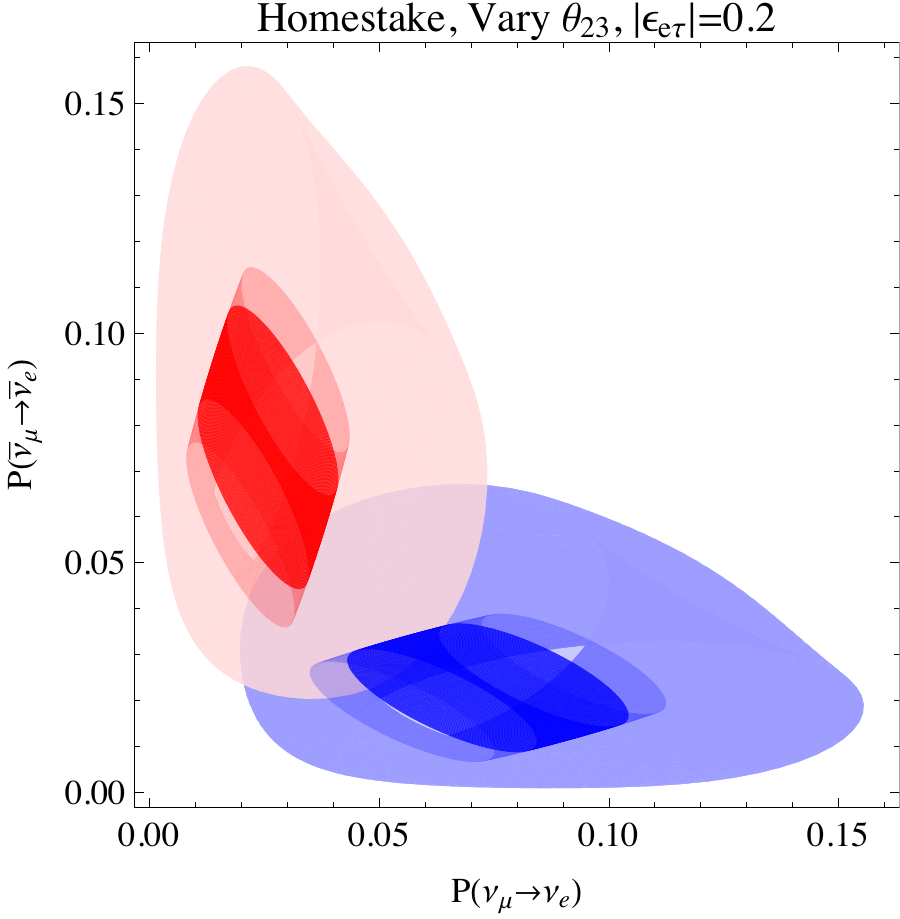}
  \caption{\emph{Top:} Here we have fixed the neutrino energy to $2~$GeV, and plotted the ensuing values of $P(\nu_{\mu}\rightarrow\nu_{e})$ and $P(\overline{\nu}_{\mu}\rightarrow \overline{\nu}_{e})$ for the Fermilab-to-Homestake baseline, 1300 km. The foreground cylinders refer to SM only interactions, for the normal (blue) and inverted (red) hierarchy, with the vacuum phase and $\theta_{23}$ varied as in Fig.~\protect{\ref{fig:NOvA23}}. The larger shaded regions come from fixing $|\varepsilon_{e\tau}| = 0.2$  and varying both the vacuum and matter phases.}
  \label{fig:LBNE}
\end{figure}

\section{Breaking NO$\nu$A Degeneracies with LBNE}
\label{sec:LBNE}
The Long-Baseline Neutrino Experiment (LBNE)~\cite{LBNE} will be a very useful tool in breaking many of the degeneracies that could easily exist after NO$\nu$A. In this section we illustrate how this can be done with the Fermilab-to-Homestake baseline (1300 km), as it yields the most sensitive probe of matter effects of the proposed LBNE baselines. 

Before considering NSI, let us first examine LBNE's SM-only sensitivity to the hierarchy.
While LBNE is not planned to run with a narrow-band beam, to facilitate comparison with our \nova~results we begin by showing the predictions in the bi-probability plane for $E_{\nu}=2$ GeV. The result is depicted in Fig.~\ref{fig:LBNE} and is to be compared with Fig.~\ref{fig:NOvA23}.

As can be seen from Fig.~\ref{fig:LBNE}, the cylinders for the two hierarchies (obtained by varying $\theta_{23}$ as in the last section) are very well separated.  Thus, under the assumption of vanishing NSI, the Homestake baseline has the capability to definitively determine the sign of the hierarchy.  
With the addition of the NSI, the overlap between the two regions is restored. Thus, for part of the parameter space, the fundamental ambiguity in determining the hierarchy remains. Yet, a measurement of, say, $(P,\overline{P}) = (0.06, 0.06)$, would actually be a very exciting development, as it would indicate the existence of NSI.  

Notice that for a significant part of the parameter space, the presence of NSI would be even more apparent. For example, notice in Fig.~\ref{fig:LBNE} that the most extreme NSI point in the normal hierarchy is more than $0.05$ away from any SM physics interpretation. Thus if LBNE can achieve a sensitivity of $P\sim0.01$ to $\nu_{e}$ appearance they could observe a near $5\sigma$ detection of nonzero NSI.  Furthermore, at this extreme point one would have some information on the magnitude of NSI since the $P(\nu_{\mu}\rightarrow\nu_{e})$ is never this large with $\varepsilon_{e\tau} =0.1$. Indeed LBNE is so sensitive to NSI that it may be able to have a $3\sigma$ signal of NSI even with $\epset =0.1$.


\begin{figure}[t]
    \includegraphics[width=0.45\textwidth]{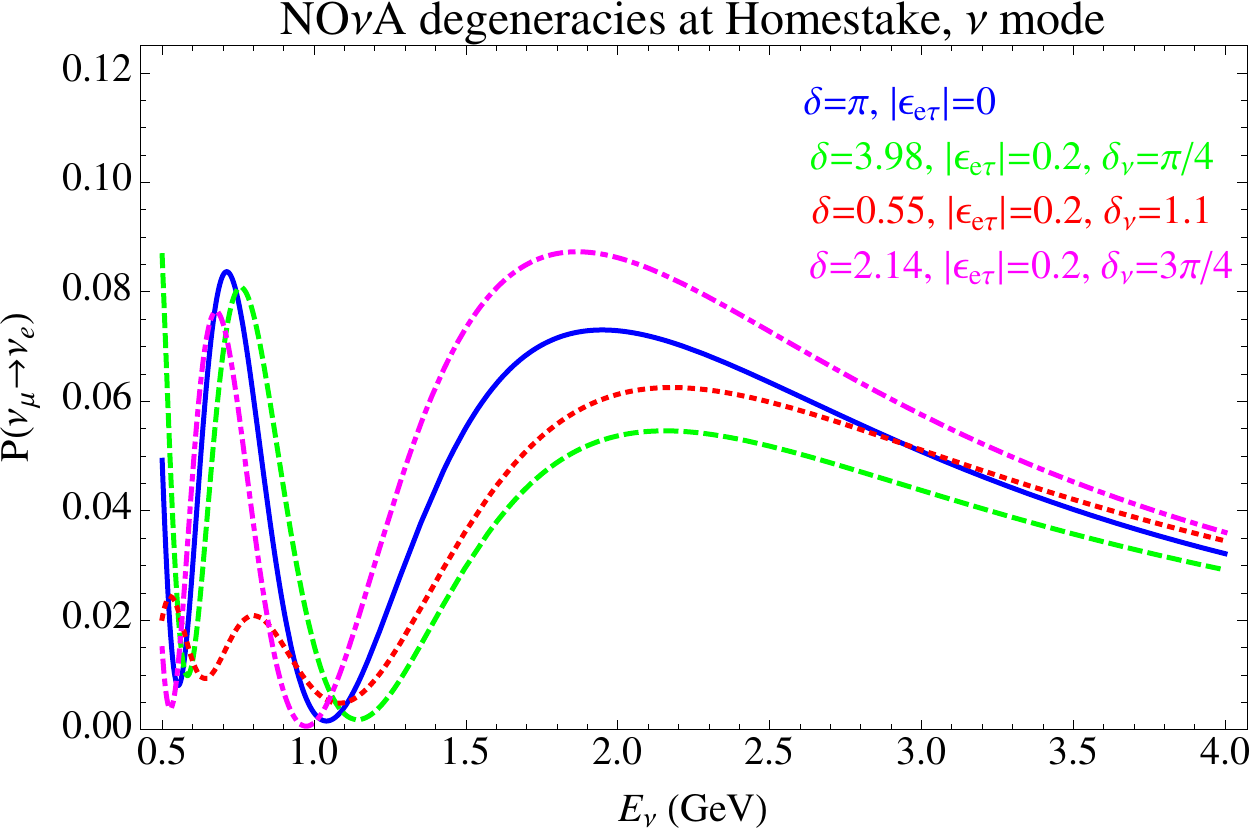}
    \includegraphics[width=0.45\textwidth]{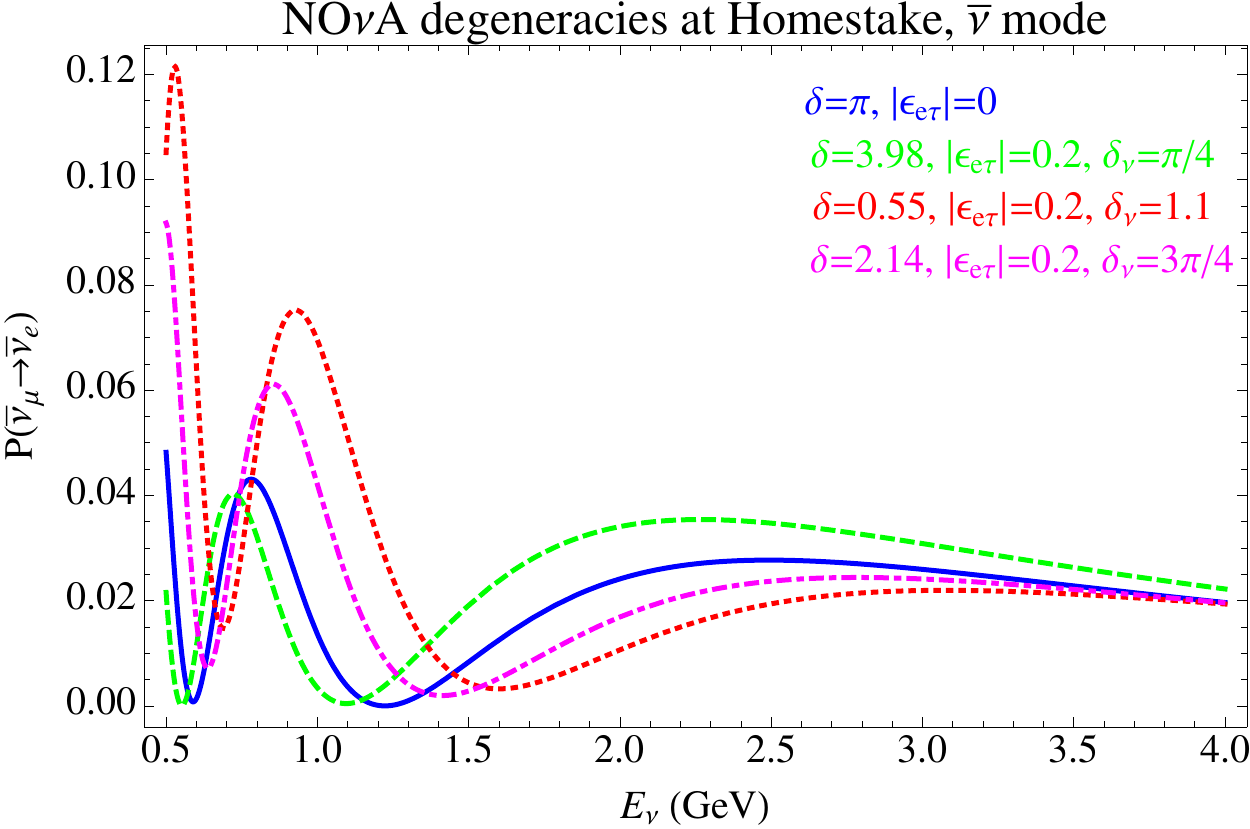}
  \caption{ An illustration how LBNE can break the NO$\nu$A degeneracies depicted in the bottom panel of Fig.~\ref{fig:NOvAbiprob2}. Here the hierarchy is normal and the color scheme is the same as in Fig.~\ref{fig:NOvAbiprob2}.}
  \label{fig:LBNEdeg}
\end{figure}

Let us now consider the full spectral information and reexamine the example of degeneracy at NO$\nu$A depicted in Fig.~\ref{fig:NOvAdeg}. Plotting the same set of curves as before, but now at the longer Homestake baseline we see, in Fig.~\ref{fig:LBNEdeg}, that the four distinct possibilities become much easier to discriminate, even with only information at 2 GeV. Beginning in the $\nu$ mode, we see that some mild ambiguity continues to exist in discerning between the red curve with $(\delta, \delta_{\nu}) = (3.98,~\pi/4)$ and the green curve with $(\delta, \delta_{\nu}) = (0.55,~1.1)$.  This is an excellent example, however, of the benefit of having broader energy sensitivity. If one has access to low-energy information, especially as low as the second oscillation maximum, then one could easily tease out the difference between the two curves.  Likewise, information in the antineutrino mode (bottom panel of Fig.~\ref{fig:NOvAdeg}) demonstrates an orthogonal method for the discrimination of the phase. Notice that in the $\overline{\nu}$ mode, the two closest curves in the $\nu$ mode are the farthest apart. %

We see that the data from \nova~and LBNE should be analyzed together. In fact, the combined analysis of data from several baselines, such as T2K, \nova, LBNE, and Deep Core promises to provide a very powerful search tool for NSI.


\section{Conclusions}
\label{sec:conc}
We have argued that present solar neutrino data may provide a hint of new interactions between matter and neutrinos.  Such interactions are not well-bounded by existing experimental probes. In this paper, we have advocated a deliberately simplified, effective framework for gauging the sensitivity of long-baseline neutrino experiments to NSI. We have further argued that the recent measurements from Daya Bay of large $\theta_{13}$ magnify the effects of NSI via quantum interference with standard vacuum oscillation physics.  We have found that near-term long-baseline experiments will be an important testing ground for the viability of NSI.  In particular, the $\nu_{\mu} \rightarrow \nu_{e}$ search by MINOS is already starting to exclude previously allowed regions of parameter space. The upcoming NO$\nu$A experiment will have increased $\nu_{\mu} \rightarrow \nu_{e}$ sensitivity, being able to see 3$\sigma$ deviations from the no-NSI hypothesis.  However, the additional $CP$-violating phase $\delta_\nu$ from NSI complicates the determination of the neutrino mass hierarchy, the octant of $\theta_{23}$ and the vacuum phase $\delta$. 
We discuss how this inherent degeneracy could be broken with information from other baselines.  To this end, we demonstrated that a long-baseline experiment with $L \sim 1300 $ km (LBNE) would be able to break many of these degeneracies and determine the hierarchy and the presence of NSI. 
Combining data from other baselines is expected to resolve the degeneracies even further. 

Lastly, we note while we only considered the impact of new physics on neutrino propagation in matter, depending on the underlying physics model of NSI, one may also encounter other new physics effects. In particular, if a new light sector is responsible for the discrepancy in the solar data, it could also play an important role in neutrino production and detection~(see for example~\cite{Nelson:2007yq}), though existing data from fixed-target experiments~\cite{Batell:2009di} and so-called ``zero distance'' oscillation experiments~\cite{Biggio:2009nt} provide strong constraints on some models of such new physics.  
We note that distinct physical effects on neutrino production, detection, and propagation could be discriminated with a long-baseline experiment if far detector data is used in conjunction with data from a near detector.


\acknowledgments
This work was supported by the LANL LDRD program. We thank the organizers of the LBNE Reconfiguration workshop (Fermilab, April 2012), the Project X Study workshop (Fermilab, June 2012) and the CETUP*  workshop (Lead, SD, July 2012) for the opportunity to present this work. A. F. also gladly acknowledges the hospitality of the Lawrence Berkeley National Laboratory, where part of this work was completed.

\bibliography{NSI}

\end{document}